Increase of the wear resistance of carbide layers deposited by Pulsed Laser Deposition in addition with an auxiliary laser

Timo Körner, Andreas Heinrich and Bernd Stritzker
Institut für Physik, Universität Augsburg, D-86135 Augsburg, Germany

Carbides stand out because of their high hardness and wear-resistance. Thus these materials are often discussed for coatings of machine tools etc.

Within this work Boron Carbide ($B_4C$) and Carbide (C) thin films were deposited on Si (100) substrates by pulsed-laser deposition technique. In order to improve the wear-resistance of the deposited films, we introduced a new working technique including the application of a second excimer laser in a special working mode. Thereby one laser was used to ablate the carbide material from a target and to deposit the material on the substrate. The light of the second laser was directed directly onto the substrate in order to modify the ablated material.

We report on details for film deposition and film properties determined by Scanning Electron Microscope, Energy Dispersive X-Ray Spectroscopy, X-Ray Diffraction, Rutherford Backscattering, Raman Spectroscopy and tribological experiments.

**I. INTRODUCTION**

Compounds containing carbon and related elements with the same or lower electronegativity especially combined with metals are so-called carbides. These materials are very important for technical and industrial applications because of their versatile properties. These are for example very high hardness (due to the covalent bonding), high wear-resistance, high melting-points, chemical resistance, low friction coefficient, large bandgap, high thermal conductivity, a low thermal expansion or a refractive index independent from wavelength.

Within this work we want to show that C and $B_4C$ can be deposited on silicon substrates by Pulsed Laser Deposition (PLD). Thereby we synthesized the carbon and boron carbide layers by a modified double laser production process, which will be described as well as the achieved film properties.

## II. EXPERIMENTAL DETAILS

For the deposition of the carbide materials a standard pulsed laser deposition system is used [1]. A KrF excimer laser (LPX300 - Lambda Physics) generates UV pulses ($\lambda$ = 248 nm) with an energy density up to 8 J/cm$^2$ on the target [2]. The Si substrate is heated up to 800°C. During the ablation process the argon hydrogen background pressure ($10^{-6} - 10^0$ mbar) and laser energy density (2-8 J/cm$^2$) is adjusted to let the plasma plume just reach the sample in a distance of about 5 cm. The stoichiometrical targets used for ablation were synthesized at Sindelhauser GmbH, Kempten, Germany.

To remove the natural SiO$_2$ layer on the silicon substrates and in order to obtain a clean and plain surface, all substrates were etched in diluted hydrofluoric acid for 30s before ablation. A cleaning with acetone and ultrasonic only, led to poorer film qualities.

The film morphology was determined by Scanning Electron Microscope (SEM) and a standard profilometer (DEKTAK). The crystallographic structure was determined by X-ray diffraction (XRD) and the stoichiometry by Energy Dispersive X-ray spectroscopy (EDX), Rutherford Backscattering (RBS), Raman Spectroscopy and finally the tribological and mechanical properties were measured.

## III. RESULTS AND DISCUSSION

*As-deposited $B_4C$ films by PLD*

In several series the growth conditions were optimized. Parameters like pulse energy density (2-8 J/cm$^2$), substrate distance (5-10 cm) and heater temperature (room temperature up to 800°C) were varied. Thereby this parameter did not influence the quality of the films dramatically. Solely the laser energy density should be above 5J/cm$^2$ due to the high ablation threshold of the carbides. As more critical the background gas has to be addressed, which is described in the following.

*Sample production within different background gases*

During the first ablations in vacuum (1·10$^{-6}$ mbar at room temperature) layers detached from the substrate. Some visible folds appeared as the film was not heated. The EDX showed the elements boron (B), graphite (C) and oxygen (O). The latter one might be due to remaining oxygen in the chamber leading to incorporation during ablation. The XRD spectra revealed no crystalline structure.

The situation did not change if a $N_2$ or a $N_2/H_2$ atmosphere was used. The surface appeared smoother, but the films still contained oxygen. Additionally an incorporation of N into the substrate could be assigned by EDX measurements.

Using $Ar/H_2$ resulted in similar surface morphologies compared with those in $N_2/H_2$. The fundamental difference was however that the argon did not diffuse into the layer. Additionally the films did not contain oxygen.

The layer thicknesses and stoichiometries were measured by RBS and revealed a homogeneous layer. To examine the textures, all samples were examined by XRD. It turned out that they were X-ray amorphous. In addition the hardness values (after Vickers) were measured to approx. 30 GPa.

In a second step the influence of a second laser beam, directed onto the substrate during the ablation should be investigated. The results are described in the following.

*Pulsed Laserdeposition of B$_4$C - modified double laser setup*

For this experiment we modified our standard laser ablation system by using a second KrF excimer (fig. 1).

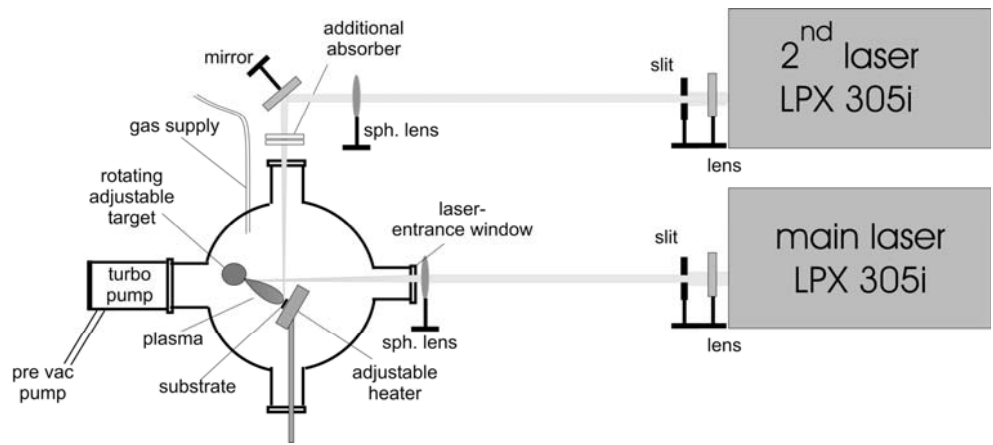

Fig. 1: Standard Setup for Pulsed Laserdeposition modified by a second KrF laser to increase the surface energy of the substrates.

The auxiliary laser light beam is directed from above on the substrate. The illuminated spot and power density desired on the substrate surface can be varied by adjusting the lenses purposefully. A further important advantage is that this system can be steered independently, whereby all parameters, e.g. energy, pulse number, focus, etc. are separately adjustable. The auxiliary laser used within this work exhibits the same specifications, as the laser used for the ablation. Thereby the laser energy had to be reduced to a minimum output energy ($<1J/cm^2$), in order to prevent a destruction of the deposited layers. Since the energy on the substrate still seemed to be too high, a special ablation mode was developed, the so-called BPM (Burst Pause Mode). That means that the total number of pulses of the ablation laser, e.g. 10000, was divided into individual bursts, e.g. 100x100 with a break of 4 seconds every 100 shots. Only during these breaks some pulses of the auxiliary laser were used.

In this case the surface contained fewer droplets and developed a wave like structure (fig. 2), which is characteristic for melted and solidified material.

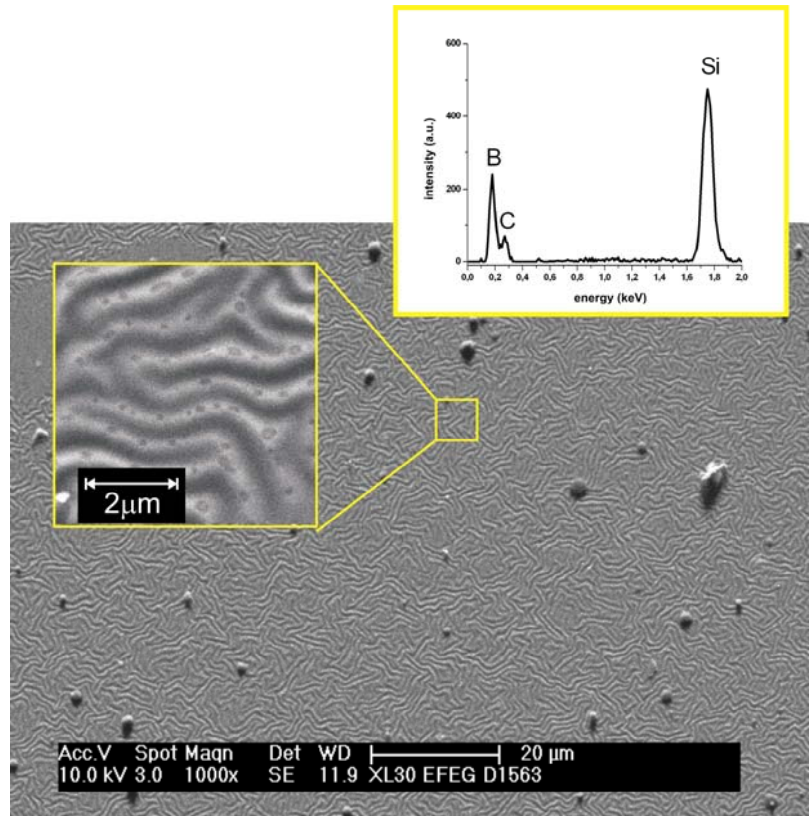

Fig. 2: ESEM picture of a 100nm thick $B_4C$ layer irradiated with 10 pulses of the auxiliary laser ($<1J/cm^2$).

The EDX analysis of the entire sample (see inset fig. 2) and the droplets on the surface clearly showed an almost stoichiometric boron carbide. The layer composition and thickness was examined by Rutherford Backscattering Spectroscopy as well. All spectra showed the correct stoichiometry and the thickness could be determined to approximately 100 nm. The hardness decreased to approx. 10 GPa. In fig. 3a the wear-resistance measurement of an as-deposited $B_4C$-layer and a $B_4C$ layer treated with the auxiliary laser is shown. For the measurement of the $B_4C$-layers a contact pressure of 1.5N was used.

The solid line represents a wear-resistance measurement of the silicon substrate. A constant friction value of 0.27 showed up after a short time. The initial fluctuations probably result from impurities on the surface, which led to the short rise to 0.35. The dotted and the dashed line represent a measurement for a B$_4$C sample deposited without and with the auxiliary laser, respectively. In both cases a high friction value can be found at the beginning. This is attributed to droplets on the substrate surface as they can be seen in figure 2. Never the less after 200 seconds this effect vanishes and the friction value clearly drops to approx. 0.03. In the case of the sample deposited with the standard PLD system the friction value turns to the substrate value after 1000s. Thus the film was ground to the substrate. In comparison to that the sample manufactured with the second laser show a clearly improved behaviour. Only after approximately four hours the friction values break in to the value of the silicon substrate. Thus with the help of the second laser, layers with clearly smaller friction value and a clearly improved wear-resistant could be manufactured.

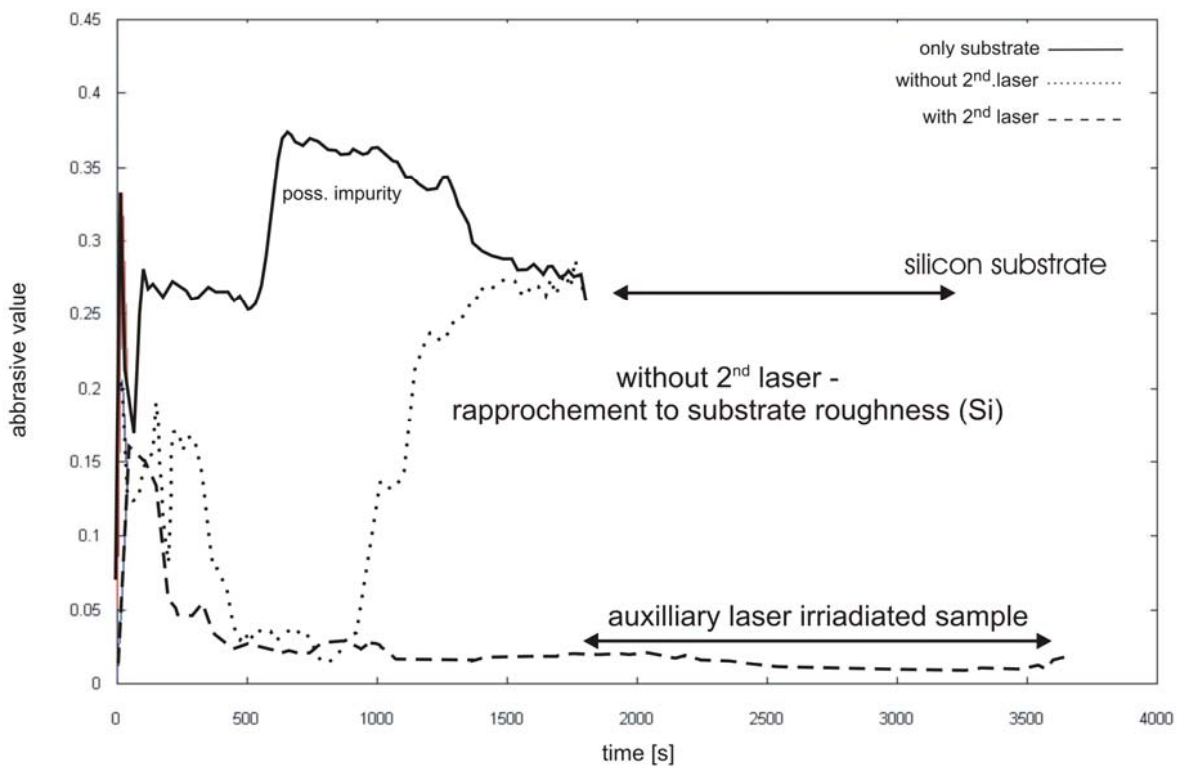

Fig. 3: Wear-resistance measurement of B$_4$C. Comparison of as-deposited with additionally illuminated areas.

*Pulsed Laserdeposition of C films*

The experiments were repeated using a C target instead of a $B_4C$ target. For the deposition of graphite layers the same parameters were used as in the previous case (argon hydrogen atmosphere, p = $5 \cdot 10^{-3}$ mbar, substrate distance 5cm, $T_H$ = 800°C). The growth rate was up to five times higher compared with $B_4C$ under the same conditions. Several samples were manufactured under the influence of the auxiliary laser in Burst Pause Mode. The samples were investigated in the same way as before.

The ESEM images of an illuminated sample with the auxiliary laser revealed the same wavy structure like in the case of $B_4C$ with few isolated droplets on it. The EDX analysis detected carbon in the layer and an oxidation of the material could be avoided by the use of $Ar/H_2$ background gas. The samples were X-ray amorphous. In total similar results were found as in the case of $B_4C$, only the thickness of the sample increased about 3 times (using the same number of pulses), due to a lower ablation threshold of the C target.

Additionally several Raman Spectroscopy measurements were accomplished. Thereby only a $sp^2$-bonded amorphous graphite layer could be determined. These values are comparable to CVD manufactured graphite layers [3]. The layer stress measurements showed a layer tension for the irradiated samples of approximately 0.9GPa. The samples treated with only one pulse of the auxiliary laser had a higher tension of 1.8GPa, falling with increasing pulse number to a value of 0.9GPa after five pulses. This means that the layer relaxed with each additional pulse.

The hardness measurements showed that the manufactured samples are very soft. No difference could be detected between the illuminated and the non-irradiated samples. The hardness was about 3GPa, which is marginally lower compared with literature values ([3]: 4.8 GPa).

The layers were finally examined on their wear-resistance. For each measurement of the carbon layers a 5.5N contact pressure was used. In fig. 4 a comparison is shown where three measurements can be specified: substrate (solid line), non-irradiated layer (dotted line) and illuminated layer (dashed line).

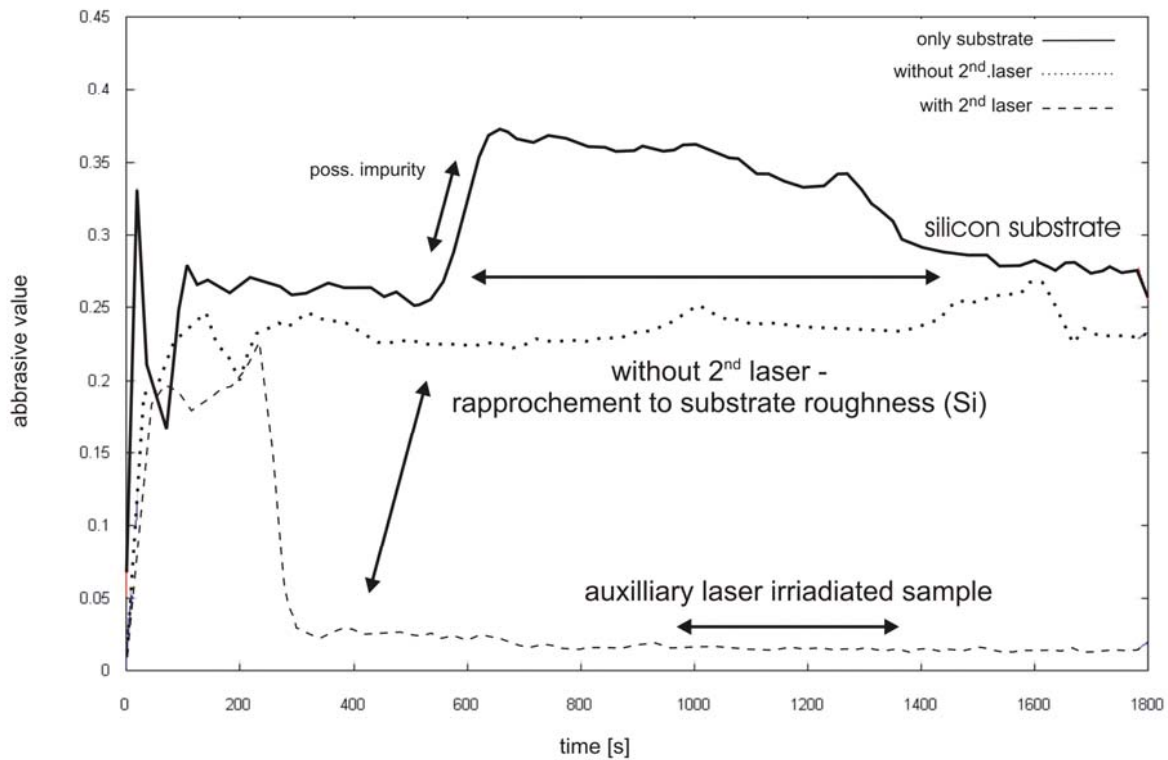

Fig. 4: Wear-resistance measurement of C. Comparison of substrate, additionally illuminated and non-irradiated sample.

For the substrate an increased friction value of approx. 0.38, which might be due to impurities, can be recognized which turns into the common substrate value of 0.27 after 1400s. In the case of the sample without auxiliary laser a reduced friction value of 0.23 can be found. As in the case of $B_4C$ the samples manufactured with the second laser show an improved behaviour. The friction value decreases to 0.02. Only after approximately eight hours the friction values breaks down to the value of the silicon substrate. An explanation for this might be that due to the laser irradiation to the film, not only the carbon, but the Si substrate as well is melted. Thus a SiC layer is formed with a clearly improved wear-resistance.

*Conclusion*

In this article, the influence of an intense UV light irradiation on a C and $B_4C$ film during deposition by PLD was reported. Using an $Ar/H_2$ background gas during deposition clearly reduced an oxidation of the deposited layers. We demonstrated that the additional energy deposited at the substrate changed its surface to a wavy structure, which is typically for melted and solidified material. Due to this the wear-resistance clearly improved in both cases.